\documentclass{PoS}
\usepackage{url}

\title{Measurement of exclusive Upsilon photoproduction in  pPb collisions at $\sqrt{s_{NN}} = 5.02$ TeV with the CMS}

\ShortTitle{Exclusive $\Upsilon$ production at CMS}

\author{\speaker{Kousik Naskar (on behalf of the CMS collaboration)}\\
	    IIT Bombay\\
        Mumbai, India\\
        E-mail: \email{kousik.naskar@cern.ch}}

\abstract{The exclusive photoproduction of $\Upsilon$(1S), $\Upsilon$(2S) and
 $\Upsilon$(3S) mesons is studied in their leptonic ($\mu^{+}\mu^{-}$) decay modes, in ultraperipheral pPb collisions
 at $\sqrt{s_{NN}} = 5.02$ TeV. The data was recorded by the CMS experiment corresponding to an integrated luminosity
 of $32.6$ nb$^{-1}$. The differential cross-section for $\Upsilon$(n) states (n=1,2,3),
 has been measured as a function of transverse momentum squared $p_{T}^{2}$, and rapidity $y$. 
The $\Upsilon$(1S) photoproduction cross-section is extracted in the region $|y|<2.2$ as a function of the 
photon-proton centre-of-mass energies $W_{\gamma p}$, in the range $91 <W_{\gamma p}< 826$ GeV.
The measurements are compared to theoretical predictions and to previous measurements.}

\FullConference{International Conference on Hard and Electromagnetic Probes of High-Energy Nuclear Collisions\\
		30 September - 5 October 2018\\
		Aix-Les-Bains, Savoie, France}

\begin{document}

\section{Introduction}
Photonuclear interactions at very high energy can be studied in 
ultraperipheral collisions (UPCs) at LHC, where protons/ions interact at large impact parameter
and therefore hadronic interactions are strongly suppressed.
The recent results of exclusive photoproduction of $\Upsilon$ and J/$\psi$ in UPCs with CMS~\cite{ref1}, 
 ALICE~\cite{ref2} and LHCb~\cite{ref3},
 reveal the importance of these measurements to probe the gluon distributions 
in nucleons and in nuclei at small Bjorken $x$, where $x$ is the fraction of target momentum carried by the gluon.
The exclusive photoproduction of vector
mesons, where a vector meson but no other
particles are produced in the event, occurs
through $\gamma p$ or $\gamma Pb$ interactions (Fig. 1a). They can be visualised 
in leading-order perturbative QCD in terms of the exchange of two gluons with no net colour transfer. Photoproduction
is strongly enhanced in heavy ions as the photon flux grows as $Z^{2}$. As the cross-section of
photoproduction of $\Upsilon$(nS) is proportional to
the square of the gluon density, it is potentially possible to probe the gluon density at small Bjorken $x$,
 which is kinematically related to the photon-proton centre-of-mass energy $W_{\gamma p}$ ($x=(M_{\Upsilon}/W_{\gamma p})^{2}$). 
If the $\Upsilon$ photoproduction is followed by the proton
breakup, the process is called "semi-exclusive" (Fig. 1b). The exchanged photon can also
 interact with a photon radiated from the other proton/ion producing an
exclusive dimuon state, which as a QED process
constitutes the main background for this analysis (Fig. 1c). 
\begin{figure}
	\begin{center}
		\hspace{-0.5cm}\includegraphics[width=50mm]{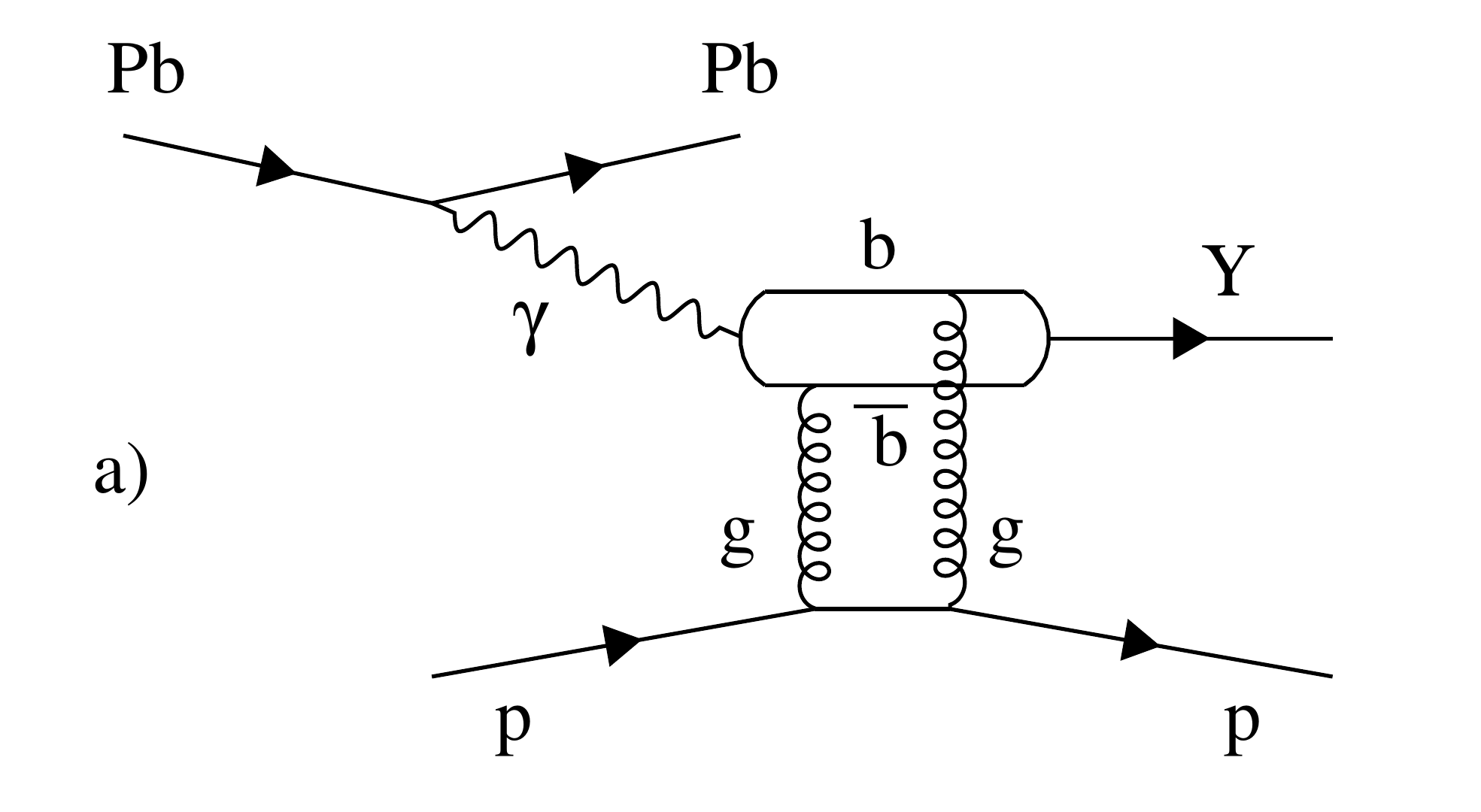}
		\hspace{-0.5cm}\includegraphics[width=50mm]{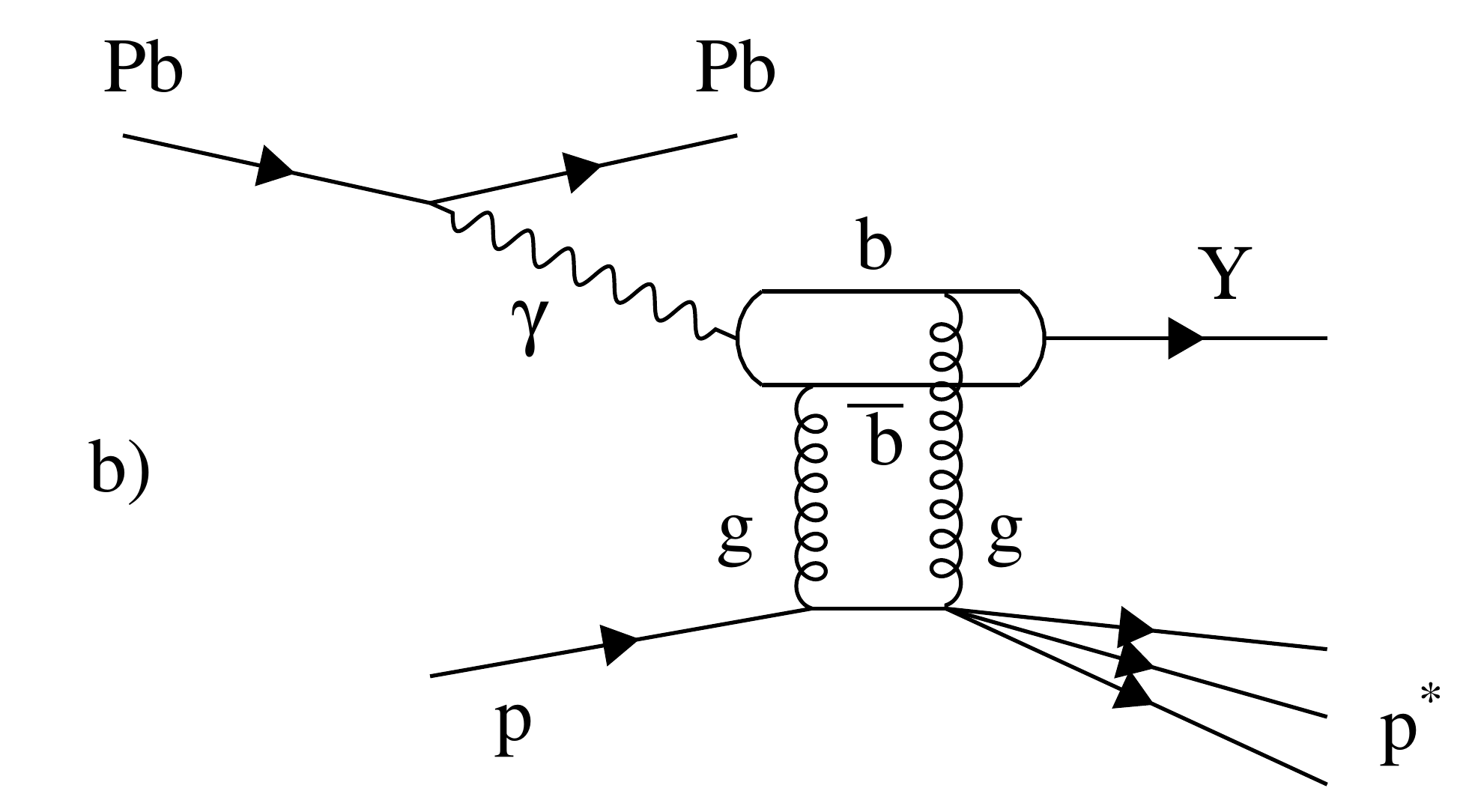}
		\hspace{-0.5cm}\includegraphics[width=50mm]{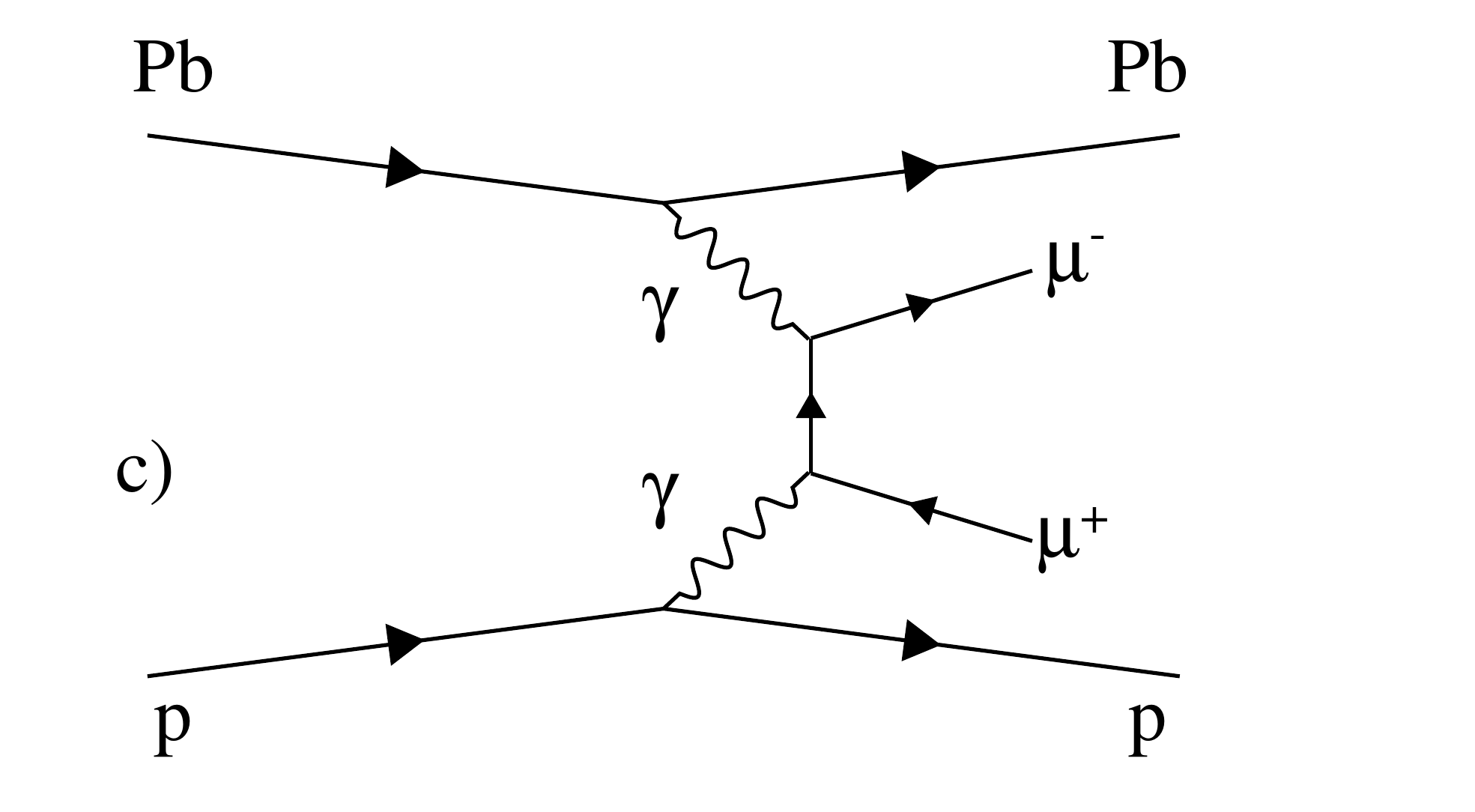}		
		\caption{Diagrams representing (a) exclusive $\Upsilon$ photoproduction, (b) proton dissociation background and (c) exclusive dimuon QED background in pPb collisions.}
		\label{fig1}
	\end{center}
\end{figure}
\section{Detector and simulation}
The CMS detector~\cite{ref4} is a general-purpose detector, having  a superconducting solenoid of $6$ m internal diameter, 
providing a magnetic field of $3.8$ T. The components of the CMS detector are a silicon pixel which sits around the LHC beampipe and
 strip tracker ($|\eta|<2.5$), a lead tungstate crystal (PbWO$_4$) electromagnetic calorimeter with rapidity coverage $|\eta|<3.0$, 
and a brass and scintillator hadron calorimeter over the range $|\eta|<3.0$, each composed of a barrel and two endcap section. Each muon station consists of several layers of aluminum drift tubes in the barrel region and cathode strip chambers
 in the endcap region, complemented by resistive plate chambers. The muon detectors are arranged in concentric cylinders around the beam line in the barrel region, and in disks perpendicular to the beam line in the endcaps ($|\eta|<2.4$). In the forward region there are several dedicated calorimeters (CASTOR, ZDC) and the TOTEM tracking detector.\\

We have used the STARLIGHT(v3.07)~\cite{ref5}
 Monte Carlo (MC) event generator to simulate exclusive
$\Upsilon$(nS) photoproduction events (Fig. 1a)
and the exclusive QED background (Fig. 1c). 
All simulated events are passed through the GEANT4-based~\cite{ref6} detector simulation and the event reconstruction chain of CMS.
\section{Event selection}
\begin{figure}
\begin{center}
	\includegraphics[width=12cm]{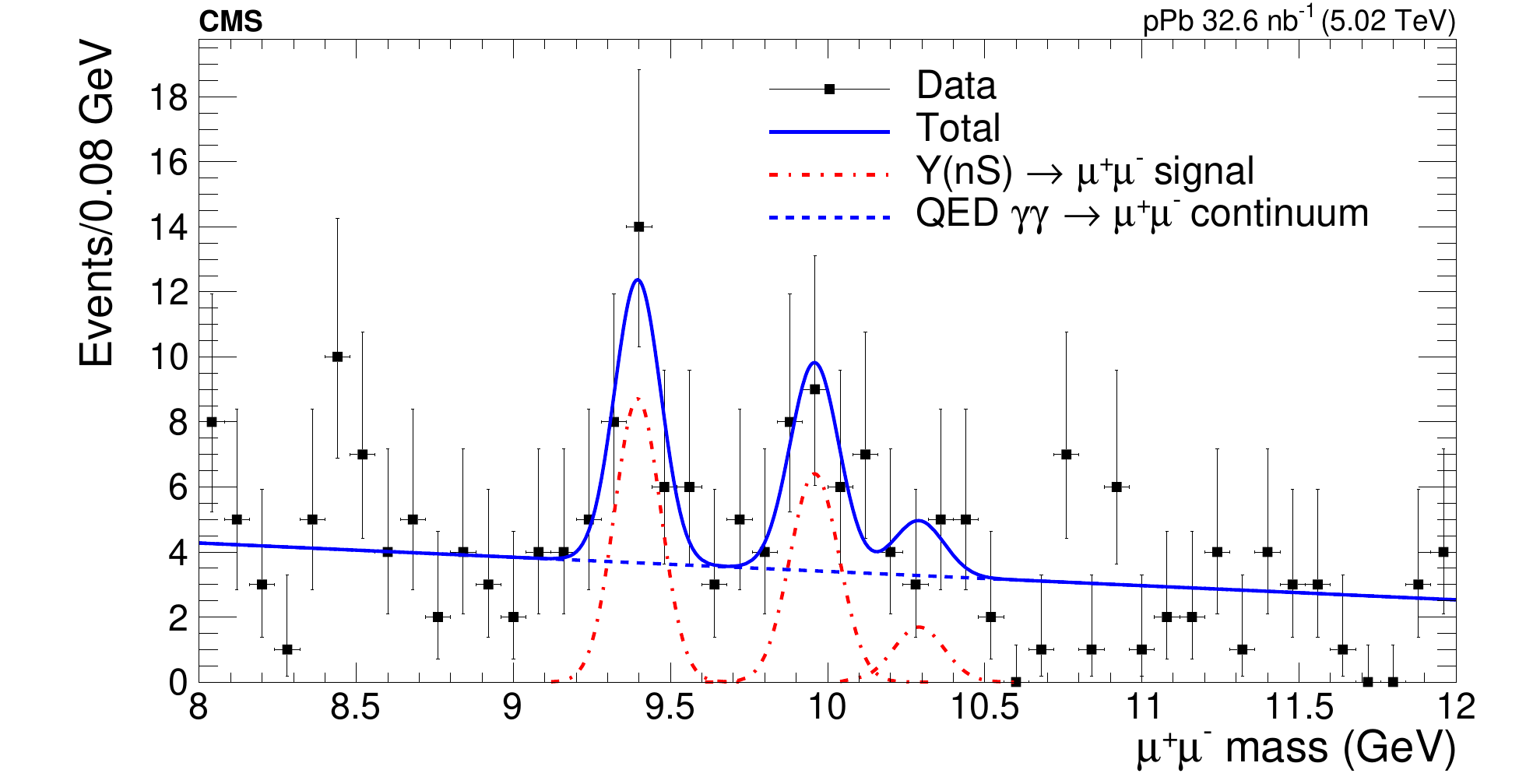}
	\caption{ Invariant  mass  distribution  of  exclusive  dimuons  in  the  range $8 < m_{\mu^{+}\mu^{-}} < 12$ GeV that pass all the event selection criteria. A linear function for the QED background (blue dashed line) plus three Gaussian distributions corresponding to the
    $\Upsilon$(1S), $\Upsilon$(2S), $\Upsilon$(3S) mesons was fitted to the data (dashed-dotted-red curves)~\cite{ref1}.}    
    \label{fig2}
\end{center}
\end{figure}

The UPC dimuon events are reconstructed at the trigger level using the High Level Trigger (HLT) algorithm, requiring at least one muon, but not more than six, tracks in the event. We have also applied the following offline muon selection criteria for the exclusive $\Upsilon$ events.
\begin{itemize}
	\item Exclusivity cut: Exclusive events are selected by requiring two
	 opposite-charge muons with a single vertex and no additional charged particles ($N_{Tracks}=2$) with track $p_{T} >0.1$ GeV.\\
	No activities in Hadronic Forward (HF) calorimeters are allowed. This is achieved by requiring leading tower energy in HF $< 5.0$ GeV, determined from detector noise distribution studies~\cite{ref1}.
	\item Single muon cut: To have good muon efficiency we have selected $p_{T}$ (single $\mu^{+}$, $\mu^{-}$) $>3.3~ \mathrm{GeV},~|\eta|<2.2$.
	\item Kinematic cuts: ($0.1 < p_{T}(\mu^{+}\mu^{-})<1.0$ GeV, $|y|<2.2$). A minimum dimuon $p_{T}$ cut is applied  to have good signal to background ratio. Also, a maximum dimuon $p_{T}$ 
cut is applied to suppress background from inclusive $\Upsilon$ and proton dissociative background.
\end{itemize}
Fig.~\ref{fig2} shows the invariant mass distribution of
$\mu^{+}\mu^{-}$ pairs in the range between $8$ and $12$ GeV that satisfy the selection criteria described above.
\section{Experimental results}
\begin{figure}
	\begin{center}
		\hspace{-0.1cm}\includegraphics[width=67mm]{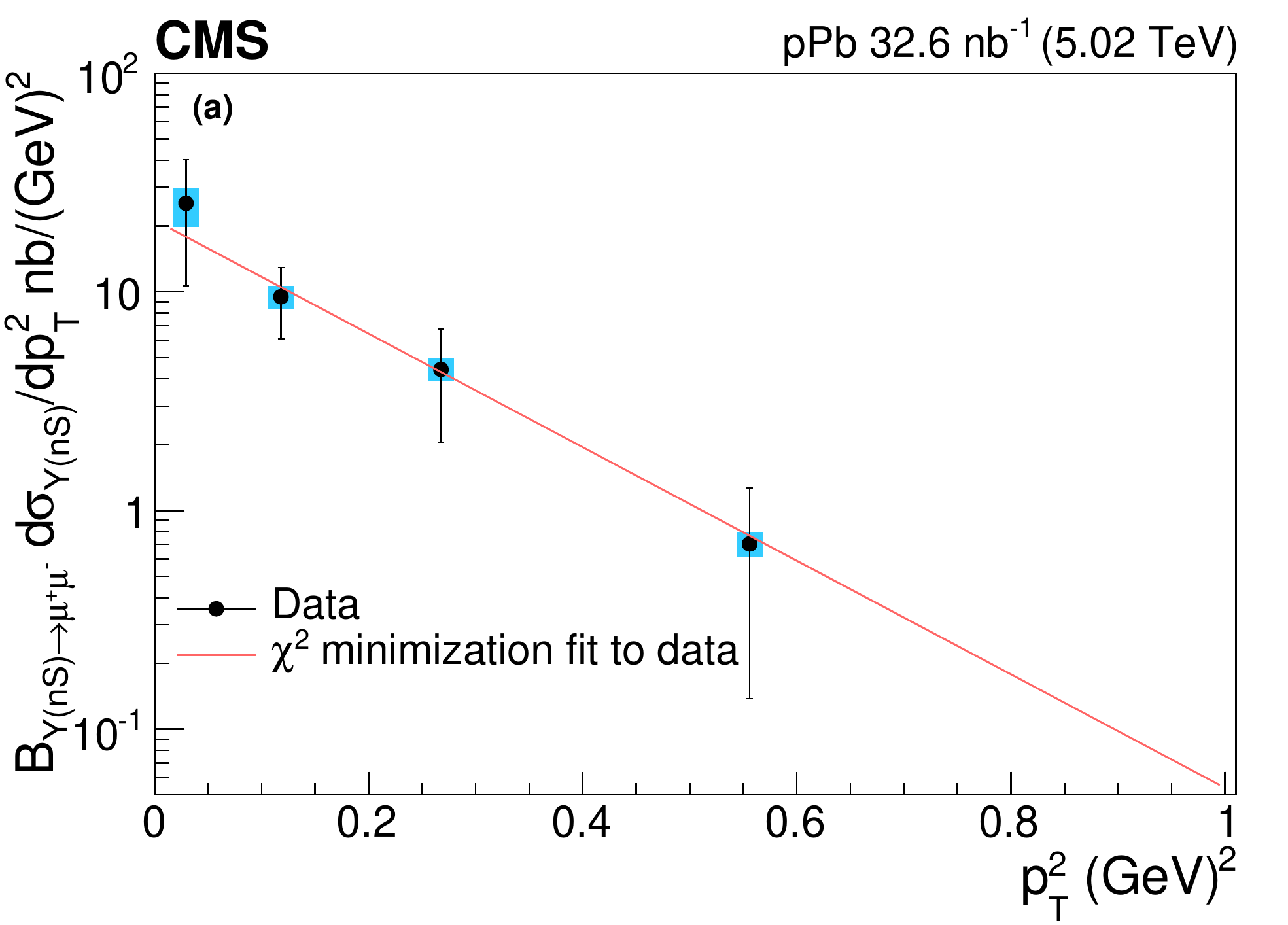}
		\hspace{0.1cm}\includegraphics[width=67mm]{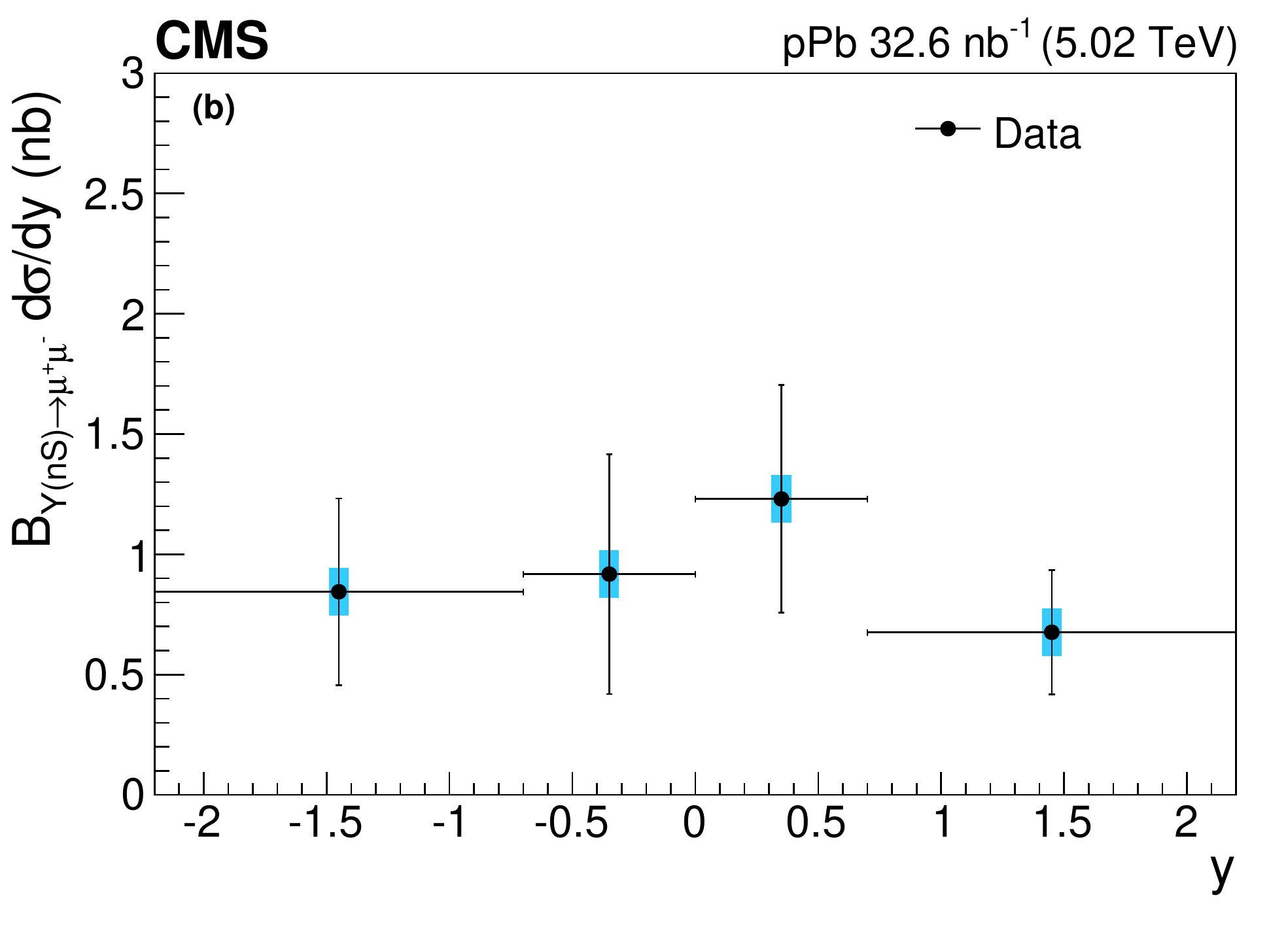}
		\caption{The differential exclusive $\Upsilon(\mbox{nS})\rightarrow  \mu^{+}\mu^{-}$ photoproduction cross-section times branching ratio, as a function of $p_{T}^{2}$ (left plot) and rapidity $y$ (right plot)~\cite{ref1}.}		
		\label{fig3}
	\end{center}
\end{figure}

The $p_T^2$ and $y$-differential cross-sections multiplied by dimuon branching fraction for exclusive $\Upsilon$(nS) photoproduction
 are extracted using the following equations
\begin{eqnarray}
\frac{d\sigma_{\Upsilon(\mbox{nS})}}{dp_{T}^{2}}\textrm{B}_{\Upsilon(\mbox{nS})\rightarrow \mu^{+}\mu^{-}} = \frac{N_{\Upsilon(\mbox{nS})}^{corr}}{\mathcal{L}\Delta p_{T}^{2}},\\
\frac{d\sigma_{\Upsilon(\mbox{nS})}}{dy}\textrm{B}_{\Upsilon(\mbox{nS})\rightarrow \mu^{+}\mu^{-}} = \frac{N_{\Upsilon(\mbox{nS})}^{corr}}{\mathcal{L}\Delta y}
\end{eqnarray}
where $N_{\Upsilon(\mbox{nS})}^{corr}$ is the background-subtracted, and acceptance-corrected number of signal events 
in the $9.1-10.6$ GeV mass region corresponding to the three $\Upsilon$ states combined in each $p_{T}^{2}$ and $y$ bin,
 ${\mathcal{L}}$ is the integrated luminosity, $\Delta p_{T}^{2}$ and $\Delta y$ are the widths of the $p_{T}^{2}$ and $y$ bins, 
and $\textrm{B}_{\Upsilon(\mbox{nS})\rightarrow \mu^{+}\mu^{-}}$ is the dimuon branching fraction. The distributions are corrected for detector resolution effects using unfolding. The differential cross-sections (multiplied by the dimuon branching fraction)  
$\Upsilon$(nS) photoproduction,
$\textrm{B}_{\Upsilon(\mbox{nS})\rightarrow \mu^{+}\mu^{-}}d\sigma_{\Upsilon(\mbox{nS})}/dp_{T}^{2}$ and 
$\textrm{B}_{\Upsilon(\mbox{nS})\rightarrow \mu^{+}\mu^{-}}d\sigma_{\Upsilon(\mbox{nS})}/dy$,
 measured in the range $|y|<2.2$, are shown in Fig.~\ref{fig3}.

The differential cross-section of exclusive $\Upsilon$(nS), $d\sigma /dp_{T}^{2}$
is fitted by the function $e^{-bp_{T}^{2}}$ (Fig.~3a). The value of the  exponential slope parameter $b = 6.0\pm 2.1$ (stat)$\pm 0.3$ (syst) GeV$^{-2}$
is extracted using a $\chi^{2}$-fit minimization method. It is in good agreement with the value $b=4.3_{-1.3}^{+2.0}$(stat) GeV$^{-2}$,
measured by the ZEUS experiment~\cite{ref7}.

\begin{figure}
	\begin{center}
		\includegraphics[width=7.7cm]{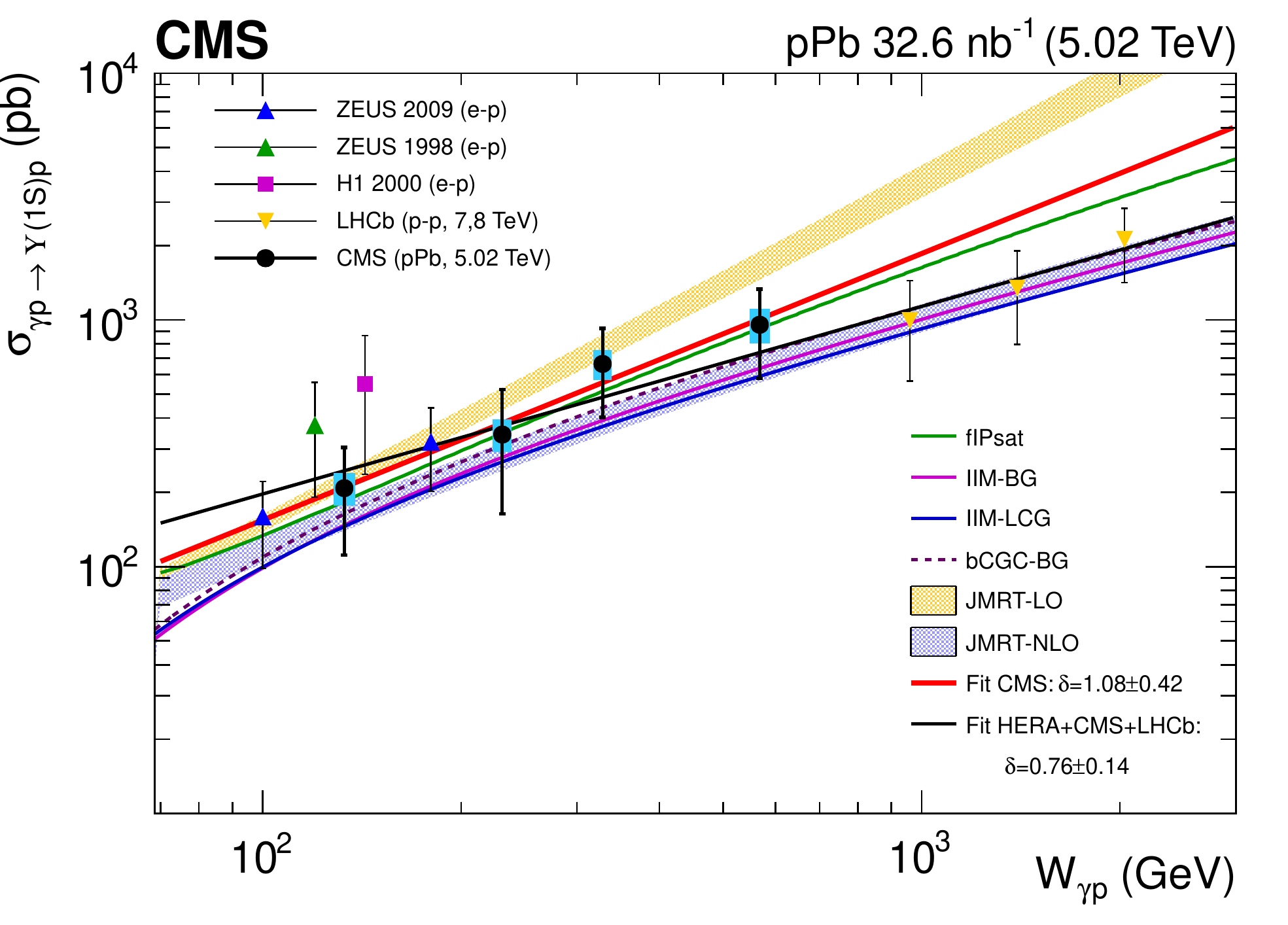}		
		\caption{ Cross section of exclusive $\Upsilon$(1S) as a function of the photon-proton centre-of-mass energy $W_{\gamma p}$~\cite{ref1}.}
		\label{fig4}
	\end{center}
\end{figure}

 The differential $\Upsilon$(1S) photoproduction cross-section is then extracted via
\begin{eqnarray}
   \frac{d\sigma_{\Upsilon(\mbox{1S})}}{dy} =\frac{f_{\Upsilon(\mbox{1S})}}{\textrm{B}_{\Upsilon(\mbox{1S})\rightarrow \mu^{+}\mu^{-}}(1+f_{FD})}[ \frac{d\sigma_{\Upsilon(\mbox{nS})}}{dy}{\textrm{B}_{\Upsilon(\mbox{nS})\rightarrow \mu^{+}\mu^{-}}}], 
\end{eqnarray}
where $f_{\Upsilon(\mbox{1S})}$ is ratio of $\Upsilon$(1S) to $\Upsilon(\mbox{1S})+\Upsilon(\mbox{2S})+\Upsilon(\mbox{3S})$ events, $f_{FD}$ is the feed-down contribution
to the $\Upsilon$(1S) from $\Upsilon(\mbox{2S})\rightarrow \Upsilon(\mbox{1S})+X$ (where $X=\pi^{+}\pi^{-}$ or $\pi^{0}\pi^{0}$) decay.
The feed-down contribution from $\chi_{b}$ states is neglected because it is a double-pomeron exchange processes.
Finally the exclusive $\Upsilon$(1S) photoproduction cross-section is measured  as a function of $W_{\gamma p}$ using the relation 
\begin{eqnarray}
\sigma_{\gamma p\rightarrow \Upsilon(\mbox{1S})p}(W_{\gamma p})=\frac{1}{\Phi}\frac{d\sigma_{\Upsilon(\mbox {1S})}}{dy},
\end{eqnarray}
in four different rapidity bins which is shown in Fig.~\ref{fig4}. The photon flux $\Phi$ is evaluated from the STARLIGHT
MC simulation.
The exclusive $\Upsilon$(1S) photoproduction cross-section is measured  in the range $91 < W_{\gamma p} < 826$ GeV, corresponds to parton fractional momenta in the proton $x \approx 10^{-4} - 10^{-2}$. This cross-section shows a power law dependence on $W_{\gamma p}^{\delta}$ and the parameter $\delta=1.08 \pm 0.42$ (CMS) is extracted from fitting. It is consistent  with the value $\delta=1.2 \pm 0.8$ obtained by ZEUS~\cite{ref8}. 
\section{Summary}  
{\tolerance=1800 We reported the first measurement of the exclusive photoproduction of $\Upsilon$(1S,2S,3S) mesons in the $\mu^{+}\mu^{-}$ decay mode for pPb collisions at $\sqrt{s_{NN}}=5.02$ TeV. The differential cross-section $d\sigma/dp_{T}^2$ and the exclusive $\Upsilon$(1S) photoproduction cross sections as a function of the photon-proton centre-of-mass energy $W_{\gamma p}$, have been measured. The present measurement provides new insight on the low-$x$ gluon distribution in the proton.\par}

\end{document}